# Wave-like Decoding of Tail-biting Spatially Coupled LDPC Codes Through Iterative Demapping

Sebastian Cammerer*, Laurent Schmalen†, Vahid Aref†, Stephan ten Brink*
*Institute of Telecommunications, Pfaffenwaldring 47, University of Stuttgart, 70569 Stuttgart, Germany
†Nokia Bell Labs, Lorenzstr. 10, 70435 Stuttgart, Germany

***Abstract*—For finite coupling lengths, terminated spatially coupled low-density parity-check (SC-LDPC) codes show a non-negligible rate-loss. In this paper, we investigate if this rate loss can be mitigated by tail-biting SC-LDPC codes in conjunction with iterative demapping of higher order modulation formats. Therefore, we examine the BP threshold of different coupled and uncoupled ensembles. A comparison between the decoding thresholds approximated by EXIT charts and the density evolution results of the coupled and uncoupled ensemble is given. We investigate the effect and potential of different labelings for such a set-up using per-bit EXIT curves, and exemplify the method for a 16-QAM system, e.g., using set partitioning labelings. A hybrid mapping is proposed, where different sub-blocks use different labelings in order to further optimize the decoding thresholds of tail-biting codes, while the computational complexity overhead through iterative demapping remains small.**

## I. Introduction

Spatially Coupled LDPC codes can reach the MAP performance under iterative belief propagation (BP) decoding by a phenomenon called *threshold saturation* [1]. However, for finite-lengths codes a non-negligible rate-loss occurs due to the required termination of the code. Tail-biting SC-LDPC codes are a way to mitigate this rate-loss. However, these codes have a worse performance than terminated SC-LDPC codes, as they have the same decoding thresholds as the underlying block LDPC codes, unless the *decoding wave* is *triggered* [2], [3].

The decoding wave can be triggered for instance if the received bits have different reliabilities. This happens in practical systems with higher-order modulation. If the pragmatic scheme of bit-interleaved coded modulation (BICM) is used, it is shown in [4], [5] and later in [6] that an optimized bit-mapping can trigger the decoding wave without termination. In particular, binary outputs of a BICM scheme can be described as a set of parallel bit channels [7]. The idea is to concentrate a specific number of bits transmitted via the better bit channel within a few sub-blocks of the tail-biting SC-LDPC code in order to trigger wave-like decoding. In the above works, the focus is BICM in combination with Gray labeling over the AWGN channel, or parallel binary erasure channels (BEC). The reliability of the different bit-channels strongly depends on the used labeling. In this work, we will investigate the potential of a different labeling in this context.

The widely used Gray labeling does not require (i.e., benefits little by) iterative demapping. However, for this very specific setup of tail-biting SC-LDPC codes a small amount of very reliable bits is required to play the effective role of the termination. Thus, a larger spread of the reliability across each bit-channel can improve the decoding threshold as long as the less reliable bits are carefully distributed over the complete code word. A large spread between the bitwise mutual information demapper functions is favorable in order to, either, minimize the number of decoding iterations, or the length of the encoding queue [6]. Therefore, a set partitioning (SP) labeling [8] is a promising candidate for further optimizations. In this paper, we investigate the advantages of iterative demapping [9] in combination with tail-biting SC-LDPC codes. It is in the nature of tail-biting codes that no rate-loss occurs. Nevertheless, the concentration of *good* bit-channels in some sub-blocks leads to a degradation of the average reliability of the remaining sub-blocks which affects the decoding thresholds. This is, in fact, the price to avoid rate-loss.

We propose a hybrid mapping scheme for tail-biting codes, where different LDPC sub-blocks use a different labeling in such a way that the decoding wave gets triggered. Even within a sub-block, different labelings are possible. A similar scheme was already proposed in [7] to match the demapping EXIT functions to the decoder function. For further optimization, we use EXIT charts to match the variable node (VND)/demapper function to the check node (CND) function. For simplicity of illustrating the method, we focus on a 16-QAM in this paper.

We show that by an iterative demapper we can improve the performance of the setup in [6]. Nevertheless a gap to capacity remains; we compare this gap with respect to the equivalent terminated ensembles.

## II. Protograph-Based SC-LDPC Codes

There are different ways to describe SC-LDPC codes [10], [11], [12]. Within this paper, we focus on protograph based codes, due to their simplicity and flexibility regarding different degree distributions. Protographs can be seen as a *blueprint* of larger graphs, where $S$ copies of the protograph are randomly connected by edge permutations. Each non-zero entry of the corresponding base matrix $\mathbf{B}$ represents the number of connected edges to this node type.

For the sake of spatial coupling, $\mathbf{B}$ can be split into sub-matrices $\mathbf{B}_i$ of dimension $M' \times N'$, i.e.

$$\mathbf{B} = \begin{bmatrix} \mathbf{B}_0 \\ \vdots \\ \mathbf{B}_{W-1} \end{bmatrix}_{WM' \times N'} .$$

In this paper we use tail-biting SC-LDPC codes [12], which do not require termination. The approach from [11] is used to construct the SC-LDPC protograph matrix $\mathbf{B}_L$, where $L$ denotes the *replication factor* (i.e., the number of sub-blocks). For a tail-biting code and a coupling window $W = 3$, we get

$$\mathbf{B}_{L,W=3,\text{tb}} = \begin{bmatrix} \mathbf{B}_0 & & & & \mathbf{B}_2 & \mathbf{B}_1 \\ \mathbf{B}_1 & \mathbf{B}_0 & & & & \mathbf{B}_2 \\ \mathbf{B}_2 & \mathbf{B}_1 & \ddots & & & \\ & \mathbf{B}_2 & \ddots & \mathbf{B}_0 & & \\ & & \ddots & \mathbf{B}_1 & \mathbf{B}_0 & \\ & & & \mathbf{B}_2 & \mathbf{B}_1 & \mathbf{B}_0 \end{bmatrix}_{LM' \times LN'}$$

and for the respective terminated code

$$\mathbf{B}_{L,W=3,\text{term}} = \begin{bmatrix} \mathbf{B}_0 & & & \\ \mathbf{B}_1 & \mathbf{B}_0 & & \\ \mathbf{B}_2 & \mathbf{B}_1 & \ddots & \\ & \mathbf{B}_2 & \ddots & \mathbf{B}_0 \\ & & \ddots & \mathbf{B}_1 \\ & & & \mathbf{B}_2 \end{bmatrix}_{(L+W-1)M' \times LN'} .$$

For a more detailed explanation of tail-biting codes, we refer the reader to [13]. The code rate for the terminated code becomes

$$R_{\text{term}} = 1 - \frac{(L+W-1)}{L} \cdot \frac{M'}{N'},$$

respectively, and for the tail-biting code

$$R_{\text{tb}} = 1 - \frac{M'}{N'} \geq R_{\text{term}}.$$

Terminated SC-LDPC codes show a superior performance compared to tail-biting codes, which basically have the same BP-thresholds as the underlying block code (see Table I). However, this excellent performance can be achieved by the tail-biting code if the BP decoding is triggered by locally improved channel reliabilities. This can be seen in Fig. 1 which illustrates the error probability of each sub-block (spatial position) within a code word under BP decoding after different decoding iterations $t$. In Fig. 1, the blocks in the spatial positions $z \in [25, 27]$ of a tail-biting ensemble are assumed to be partially known, for instance using shortening or transmission over a reliable side-channel.

It was shown that successful BP-decoding is possible close to the MAP threshold, if we carefully interleave the code bits of different spatial positions between channels [4], [5], [6]. In particular, we can seed an effective termination for threshold saturation by more channel uses of the better channel at some specific positions.

## III. System-Model

In order to minimize the rate-loss, we investigate tail-biting SC-LDPC codes with iterative demapping. It was shown in [13] that SC-LDPC codes exhibit an approximately *universal* behavior in combination with iterative demapping, as long as the code is carefully designed. This means that for a non-Gray labeling the difference in decoding performance remains small [13, Fig. 3]. Nevertheless, the price to pay is an increased number of decoding iterations and the extra computational complexity of iterative demapping. Therefore, we propose a hybrid mapping, which uses the SP labeling only for a few number of sub-blocks $T$. All other $L - T$ sub-blocks use a Gray labeling, without iterative demapping.

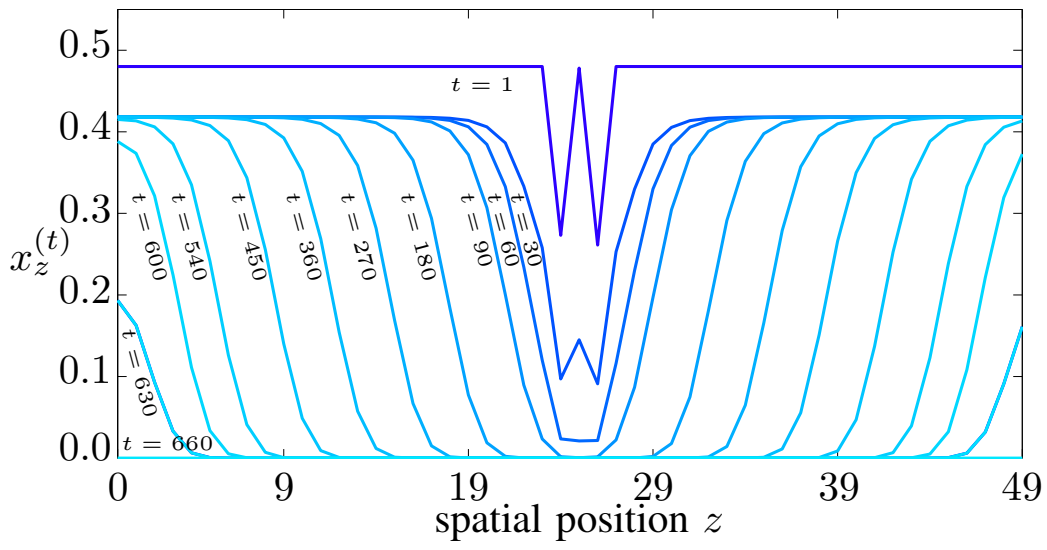

Figure 1. Wave-like decoding (density evolution) of a tail-biting $(3, 6, 50, 3)$ SC-LDPC ensemble under BP decoding over the erasure channel. Decoding is triggered by a fraction of known bits at spatial positions $z \in [25, 27]$.

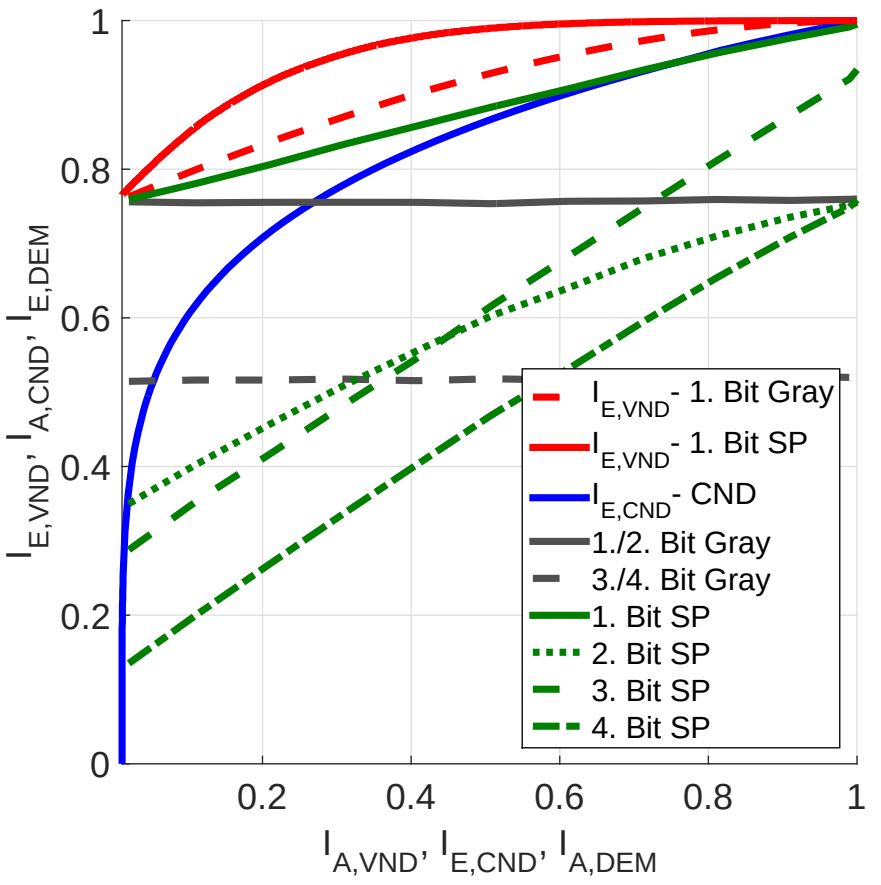

Figure 2. Bitwise EXIT functions of iterative demappers with Gray and set partitioning labeling.

In order to investigate the influence of the labeling, we extend the block diagram of [13] by the possibility of having two different labelings, as depicted in Fig. 3. We select a 16-QAM for our system setup, since it allows the usage of "true" 2-D labelings that cannot be decomposed into two independent labelings per I-/Q-channel, respectively, i.e., the Euclidean distance between two signal points can be optimized with two degrees of freedom. However, the price to pay is a 2-D demapper rather than two simple 1-D demappers. Among many possible labelings [7], a promising labeling is Ungerboeck's set partitioning (SP) labeling [8]. Fig. 2 shows the conventional EXIT curve of the demapper of this labeling, averaged over *all* bits, as well as per-bit EXIT curves describing the contribution of the *individual* bits of the labeling; obviously, the best bit-channel from SP outperforms the two best channels from Gray labeling. This can be used to improve the VND curve and allows higher decoding thresholds.

Let $K$ denote the number of different bit-channels of a given constellation, e.g., for a 16-QAM $K = \log_2(16) = 4$. Additionally, $D_{i,j}$ denotes the fraction of bits of spatial positions $j$ $(1 \leq j \leq L)$ transmitted via bit-channel $i$, where $1 \leq i \leq K$ denotes the bit-channels using Gray labeling and $K+1 \leq i \leq 2K$ the bit-channels with SP labeling. Thus, $\mathbf{D}$ defines a $2K \times LN'$ matrix containing the fractions of bits transmitted through a specific bit-channel for each of the $LN'$ spatial positions . One can easily infer the following constraints

$$\sum_i D_{i,j} = 1;\ \forall j \tag{1}$$

$$\begin{aligned}\sum_j D_{1,j} = \ldots = \sum_j D_{K,j}\\ \sum_j D_{K+1,j} = \cdots = \sum_j D_{2\cdot K,j}\end{aligned} \tag{2}$$

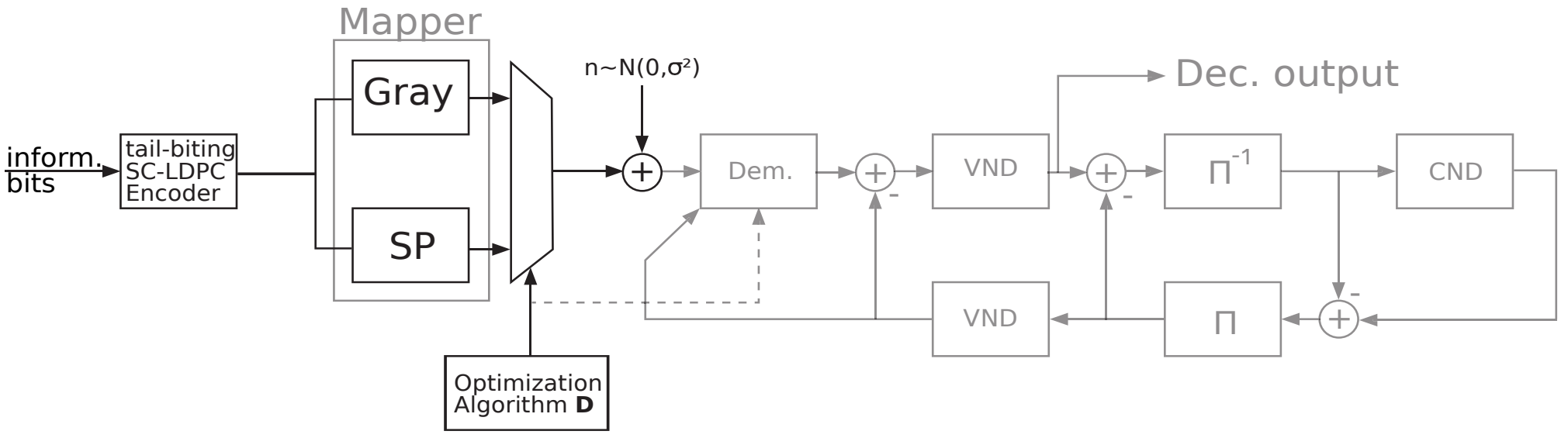

Figure 3. Block diagram of iterative demapping in combination with iterative BP LDPC decoding, based on Fig. 2 in [13].

### A. Protograph Density Evolution for Iterative Demapping

We apply density evolution for SC-LDPC codes using a Gaussian approximation (GA) including iterative demapper throughout this paper. Hence, the density evolution only tracks the mean values of the messages passed within the decoder/demapper. It was already observed in [6] that this approximation is sufficient and only a negligible difference in the thresholds occurs. Hence, we expect no further insight by applying full density evolution.

Let $\mu_{i\leftarrow j}$ denote the mean value passing from CND $c_j$ to its adjacent VND $v_i$ and let $\mu_{i\rightarrow j}$ denote the mean value passing from VND $v_i$ to CND $c_j$. The update rules become [13], [14]

$$\mu_{i\leftarrow j} = \phi^{-1} \left(1 - \left[1-\phi\left(\mu_{i\rightarrow j}\right)\right]^{B_{i,j}-1} \prod_{k=1;k\neq i}^{LN'} \left[1-\phi\left(\mu_{i\rightarrow j}\right)\right]^{B_{k,j}}\right),$$

with $\phi(\mu)$ as in [14]

$$\phi(x) = \begin{cases} 1 - \frac{1}{\sqrt{4\pi x}} \int_{-\infty}^{\infty} \tanh\left(\frac{u}{2}\right) \exp\left(-\frac{(u-x)^2}{4x}\right) \mathrm{d}u, & x > 0 \\ 1, & x = 0 \end{cases}$$

The variable node update from VND $v_i$ to CND $c_j$ is

$$\mu_{i\rightarrow j} = \mu_{D\rightarrow i} + (B_{ij}-1)\cdot \mu_{i\leftarrow j} + \sum_{k=1,k\neq j}^{LM'} B_{ik}\cdot \mu_{i\leftarrow k}.$$

Using the bit-wise demapper EXIT function $f_{D,i}(I_{A,\mathrm{DEM}})$ of bit channel $i$, the mean value of messages from demapper to variable node becomes

$$\mu_{D\rightarrow i} = J^{-1}\left(\sum_{l=1}^{2K} D_{l,i}\cdot f_{D,l}\left(J\left(\sum_{k=1}^{LM'} B_{ik}\cdot \mu_{i\leftarrow k}\right)\right)\right).$$

We use the approximation [7]

$$J(\mu) \approx \left(1 - 2^{-H_1\cdot(2\mu)^{H_2}}\right)^{H_3}$$

and

$$J^{-1}(I) \approx \frac{1}{2}\left(-\frac{1}{H_1}\cdot \log_2\left(1 - I^{\frac{1}{H_3}}\right)\right)^{\frac{1}{H_2}}$$

with $H_1 = 0.3073$, $H_2 = 0.8935$ and $H_3 = 1.1064$.

For a more detailed explanation we refer the interested reader to [13], [14], [15].

### B. EXIT-Chart Approximation of Thresholds

It is a well-known fact that terminated SC-LDPC codes approach the MAP performance of the underlying block LDPC code under BP decoding, when $L$ and $W$ are sufficiently large [1], [10]. The MAP threshold of a block LDPC code used for transmission over a binary erasure channel (BEC) can be approximated by EXIT charts [3], which do not require computational extensive simulations. In what follows, we investigate if we can use this simple approximation to predict the performance of protograph-based SC-LDPC codes with iterative demapping.

Assuming irregular degree distributions $\lambda(Z)$ and $\rho(Z)$ and Gaussian distributed decoder messages, the extrinsic mutual information (MI) of the check node update $I_{e,c}$ can be approximated as

$$I_{e,c} \approx 1 - \sum_{j=2}^{c_{max}} \rho_j \cdot J\left((j-1)\cdot J^{-1}\left(1 - I_{a,c}\right)\right).$$

The extrinsic MI of the variable node update $I_{e,v}$ is

$$I_{e,v} = \sum_{i=2}^{v_{max}} \lambda_i \cdot J\left(\mu_c + (i-1)\cdot J^{-1}\left(I_{a,v}\right)\right), \tag{3}$$

with [13], [16]

$$\mu_c = J^{-1}\left(f_D\left(\sum_{i=1}^{v_{max}} L_i \cdot J\left(i\cdot J^{-1}\left(I_{a,v}\right)\right)\right)\right)$$

where $f_D(I_{a,\text{DEM}})$ describes the demapper function [7] of the used mapping and $L(Z)$ the normalized variable node degree distribution from node perspective.

The BEC MAP threshold[1] $\tilde{\gamma}^*_{EXIT}$ of the uncoupled random ensemble with given degree distribution is defined to be the $E_b/N_0$ value for which the area $A_{1,2}$ between the first and second intersection of the VND and CND curve and the area $A_{2,3}$ between the second and third intersection of the same curves become equal [3], as showcased in Fig. 4. Note that the CND curve is not affected by the demapper, while the VND curve depends on the demapper according to (3). The investigated tail-biting codes are

- Code $C_1$ - regular $(3, 6)$ LDPC code with $R_{\text{tb}} = 0.5$ and $W = 3$
- Code $C_2$ - regular $(4, 8)$ LDPC code with $R_{\text{tb}} = 0.5$ and $W = 4$
- Code $C_3$ - constructed from regular protograph $\mathbf{B}_0 = \mathbf{B}_1 = (2\,2\,2\,2)$ with $R_{\text{tb}} = 0.75$ and $W = 2$
- Code $C_4$ - constructed from irregular protograph $\mathbf{B}_0 = (1\,2\,1\,2)$ and $\mathbf{B}_1 = (3\,2\,3\,2)$ with $R_{\text{tb}} = 0.75$ and $W = 2$

Table I shows the results from the EXIT chart prediction for the codes used throughout this paper in comparison to the density evolution results [17] for the terminated coupled ensemble. For the uncoupled case, the well-known EXIT chart thresholds (*open decoding tunnel*) approximate the conventional BP decoding thresholds. The predicted EXIT-chart-based coupled thresholds $\tilde{\gamma}^*_{EXIT}$ coincidence with the approximated BP decoding thresholds $\gamma^*_{BP,\text{sc}}$ of the terminated coupled ensemble for most of the codes, as expected. We need to emphasize that $\gamma^*_{BP,\text{sc}}$ in general does not necessary equal $\tilde{\gamma}^*_{EXIT}$ (especially for finite $W$ and $L$), e.g., for $C_3$ and SP, the thresholds differ strongly.

For all investigated uncoupled LDPC codes, Gray labeling outperforms SP labeling under BP decoding, however the performance depends on the specific degree profile and protograph structure. The codes $C_1$ and $C_2$ exemplify that a good $\gamma^*_{BP,\text{un}}$ does not always result in a desirable $\gamma^*_{BP,\text{sc}}$ and vice versa. It was observed in [13] that $C_4$ shows an approximately universal behavior with respect to the labeling and is therefore a promising candidate for a hybrid mapping scheme.

$C_3$ and $C_4$ have the same EXIT thresholds, because they have the same degree profile[2]. Nevertheless, $C_4$ shows a stronger (one-sided) termination and thus a one-sided decoding wave, whereas $C_3$ does not properly trigger wave-like decoding for SP mapping. This explains the differences between EXIT and DE results and is explained in [13].

[1]Remark: Strictly speaking, the applied area theorem is only valid for the BEC and randomly coupled ensembles. Nevertheless, we expect that it is a good match to approximate the MAP threshold of related ensembles over other channels as well.

[2]Remark: Since these codes are protograph based ensembles, the EXIT chart does not directly apply.

[3]Remark: for all codes $L = 50$, however, for a better comparison the rate loss due to termination effects is not considered here.

Table I
DECODING THRESHOLDS OF DIFFERENT CODES AND LABELINGS FOUND BY EXIT CHARTS AND DENSITY EVOLUTION FOR 16-QAM OVER THE AWGN CHANNEL

| | | uncoupled ($\gamma^*_{BP,\text{un}}$) | | | coupled[3] | |
|---|---|---|---|---|---|---|
| | | Gray | SP | | Gray | SP |
| $C_1$ | EXIT | 3.412 dB | 4.705 dB | EXIT ($\tilde{\gamma}^*_{EXIT}$) | 2.582 dB | 2.189 dB |
| | DE | 3.412 dB | 4.741 dB | DE ($\gamma^*_{BP,\text{sc}}$) | 2.545 dB | 2.138 dB |
| $C_2$ | EXIT | 4.000 dB | 5.667 dB | EXIT ($\tilde{\gamma}^*_{EXIT}$) | 2.352 dB | 2.108 dB |
| | DE | 4.005 dB | 5.702 dB | DE ($\gamma^*_{BP,\text{sc}}$) | 2.279 dB | 2.054 dB |
| $C_3$ | EXIT | 5.460 dB | 6.477 dB | EXIT ($\tilde{\gamma}^*_{EXIT}$) | 4.753 dB | **4.556 dB** |
| | DE | 5.471 dB | 6.501 dB | DE ($\gamma^*_{BP,\text{sc}}$) | 4.770 dB | **5.134 dB** |
| $C_4$ | EXIT | 5.460 dB | 6.477 dB | EXIT ($\tilde{\gamma}^*_{EXIT}$) | 4.753 dB | 4.556 dB |
| | DE | 5.471 dB | 6.501 dB | DE ($\gamma^*_{BP,\text{sc}}$) | 4.753 dB | 4.678 dB |

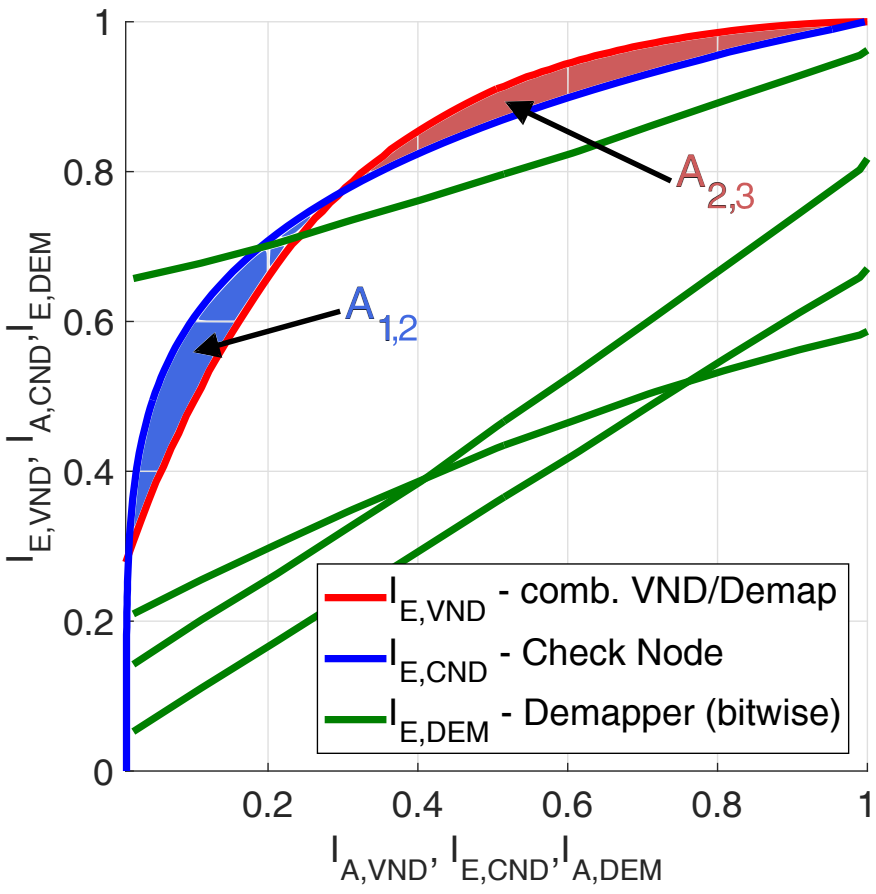

Figure 4. EXIT-Chart of a regular (3,6) block LDPC ensemble with iterative demapping and SP labeling, the areas $A_{1,2}$ and $A_{2,3}$ between the intersections are equal.

## IV. Optimization of the Mapping

Instead of shortening bits (terminated SC-LDPC) we consider tail-biting codes in combination with an optimized mapper, i.e., bits from the better bit channels are concentrated on several sub-blocks of the overall codeword [4], [5], [6].

As we can see in Table I, $\gamma^*_{BP,\text{un}}$ does not improve with SP labeling for the selected codes. Nevertheless, when we pick bits from the better bit channels, SP in conjunction with iterative demodulation enables a more reliable bit channel (see Fig. 2). However, the drawback is that the remaining other bit channels become worse in comparison to the Gray labeling. Although, $\gamma^*_{BP,\text{sc}}$ can improve in general for SP labeling, we propose a hybrid scheme, where only the first blocks are using iterative demapping. The advantage is that iterative demapping does only effect the decoding complexity at the first sub-blocks.

In Fig. 5, we plot the decoding thresholds of the different tail-biting codes as a function of the number of sub-blocks $T$ transmitted via the best available bit channel (as a result of an optimized mapper). The three curves per code show the optimization results for Gray, SP and a hybrid labeling, whereas the hybrid labeling uses SP labeling only for $T$ sub-blocks (in combination with an optimized mapping scheme). The idea is that for large enough $T$, the BP algorithm can locally decode the more reliable code bits and thus, this region can be considered as an effective termination. Obviously, the decoding performance of the remaining $L-T$ sub-blocks degrades, because transmission takes mainly place over the remaining (worse) bit channels.

The simulation results show that the number of required sub-blocks $T$ to trigger wave-like decoding decreases in comparison to Gray labeling, if SP labeling is involved, i.e., less bits from the good channel need to be concentrated on a specific spatial position. For comparison with our previous results in [6], we use the same code $C_1$, i.e., a regular tail-biting $(v=3, c=6, L=50, W=3)$ SC-LDPC code. Nevertheless, these optimizations require efficient (sub-optimal) algorithms, such as differential evolution. Each single spatial position $j$ of $\mathbf{D}$ defines additional degrees of freedom for this optimization; it turns out that the optimization can be relaxed without a significant loss of performance by a *uniform distribution* [6] with only two different vectors for $T_{uni}$ and $L-T_{uni}$ spatial positions.

Interestingly, the optimized tail-biting ensemble $C_1$ with SP labeling even outperforms the terminated SC-LDPC ensemble (Gray labeling) as the corresponding $\gamma^*_{BP,\mathrm{sc}}$ is better[4]. The hybrid mapping (here: $T$ blocks with SP labeling and $L-T$ blocks with Gray labeling) avoids iterative demapping within the remaining $L-T \gg T$ blocks, but also decreases the thresholds a bit. Nonetheless, the hybrid mapping still outperforms the terminated SC-LDPC with Gray labeling, since no rate loss (because of termination effects) occurs.

For $C_3$ and $C_4$ we observe almost the same thresholds as the terminated versions, but again, no rate loss occurs, i.e., the code rate is larger. Thus, the remaining gap to capacity effectively decreases. As discussed earlier, a difference between the two codes can be seen for the optimized SP labeling, since $C_3$ shows a lack of performance for SP labeling.

Hybrid mappings turn out to be the most promising scheme, since the required number of decoding iterations is mainly defined by the Gray labeling and, thus, smaller than for SP labeling.

As an amendment to our previous work [6], one may notice that a smaller encoding buffer is required when using SP or hybrid labeling, which becomes important for practical, latency-constrained, implementations of such systems.

## V. Conclusion

In this paper, we have investigated the use of hybrid iterative demodulation to improve the decoding performance of tailbiting spatially coupled LDPC codes. We have found that the hybrid mapping in conjunction with tailbiting codes allows gains that can even outperform terminated codes, without having an inherent rate loss.

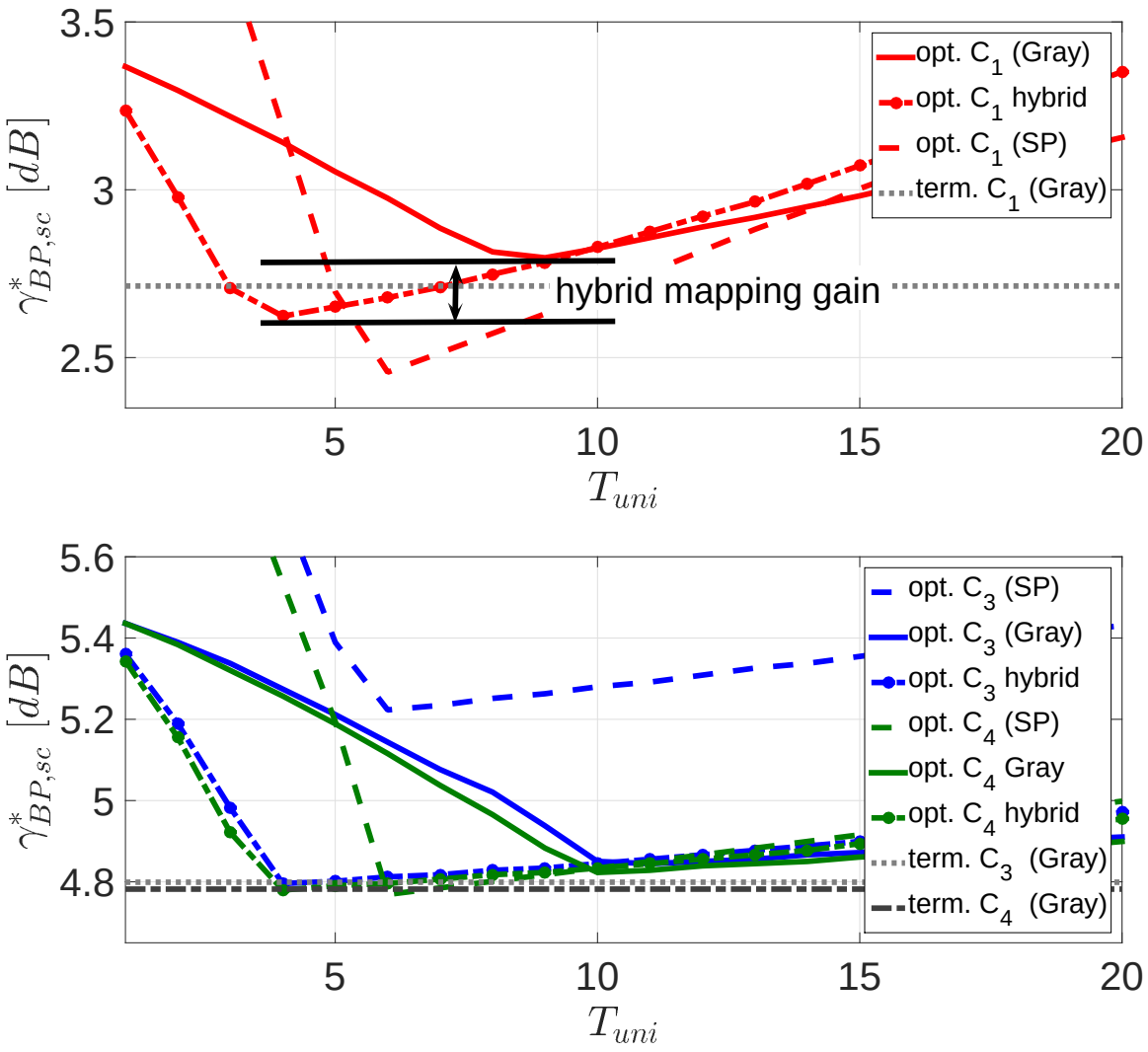

Figure 5. Optimized decoding threshold of tail-biting ensembles for several labelings and mapping schemes found by density evolution. For all codes $L=50$. Remark: A rate loss occurs for the terminated codes.

## References

[1] S. Kudekar, T. J. Richardson, and R. L. Urbanke, "Threshold saturation via spatial coupling: Why convolutional LDPC ensembles perform so well over the bec," *IEEE Trans. Inform. Theory*, vol. 57, Feb 2011.

[2] S. Kudekar, C. Méasson, T. J. Richardson, and R. L. Urbanke, "Threshold saturation on BMS channels via spatial coupling," in *Proc. Internat. Symp. Turbo Codes*, vol. 59, 2010, pp. 309–313.

[3] S. Kudekar, T. J. Richardson, and R. L. Urbanke, "Wave-like solutions of general 1-d spatially coupled systems," *IEEE Trans. Inf. Theory*, vol. 61, no. 8, pp. 4117–4157, Aug 2015.

[4] C. Häger, A. Graell i Amat, A. Alvarado, F. Brännström, and E. Agrell, "Optimized bit mappings for spatially coupled LDPC codes over parallel binary erasure channels," *Proc. IEEE Int. Conf. on Commun. (ICC)*, pp. 2064–2069, June 2014.

[5] C. Häger, A. Graell i Amat, F. Brännström, A. Alvarado, and E. Agrell, "Terminated and tailbiting spatially coupled codes with optimized bit mappings for spectrally efficient fiber-optical systems," *Journal of Lightwave Technology*, vol. 33, no. 7, pp. 1275–1285, 2015.

[6] S. Cammerer, V. Aref, L. Schmalen, and S. ten Brink, "Triggering wave-like convergence of tail-biting spatially coupled LDPC codes," *2016 Annual Conference on Information Science and Systems (CISS)*, pp. 93–98, March 2016.

[7] F. Schreckenbach, "Iterative decoding of bit-interleaved coded modulation," *Ph.D. dissertation, Technical University Munich (TUM)*, 2007.

[8] G. Ungerboeck, "Channel coding with multilevel/phase signals," *IEEE Trans. Inform. Theory*, vol. 28, no. 1, pp. 55–67, Jan. 1982.

[9] S. ten Brink, J. Speidel, and R.-H. Yan, "Iterative demapping and decoding for multilevel modulation," *Proc. IEEE Globecom Conf.*, 1998.

[10] S. Kudekar, T. Richardson, and R. Urbanke, "Spatially coupled ensembles universally achieve capacity under belief propagation," *IEEE Trans. Inform. Theory*, vol. 59, no. 12, pp. 7761–7813, 2013.

[11] D. G. M. Mitchell, M. Lentmaier, and D. J. Costello, "Spatially coupled LDPC codes constructed from protographs," *IEEE Trans. Inf. Theory*, vol. 61, no. 9, pp. 4866–4889, Sept 2015.

[12] M. B. Tavares, K. S. Zigangirov, and G. P. Fettweis, "Tail-biting LDPC convolutional codes," in *Proc. IEEE Int. Symp. on Inform. Theory*, June 2007, pp. 2341–2345.

[13] L. Schmalen and S. ten Brink, "Combining spatially coupled LDPC codes with modulation and detection," in *Proc. ITG SCC*, January 2013.

[14] S.-Y. Chung, T. J. Richardson, and R. L. Urbanke, "Analysis of sum-product decoding of low-density parity-check codes using a Gaussian approximation," *IEEE Trans. Inform. Theory*, vol. 47, Feb 2001.

[15] T. Richardson and R. Urbanke, *Modern Coding Theory*. Cambridge University Press, 2008.

[16] S. ten Brink, G. Kramer, and A. Ashikhmin, "Design of low-density parity-check codes for modulation and detection," *IEEE Trans. Commun.*, vol. 52, no. 4, pp. 670–678, April 2004.

[17] S. Jayasooriya, M. Shirvanimoghaddam, L. Ong, G. Lechner, and S. J. Johnson, "New density evolution approximation for LDPC and multi-edge type LDPC codes," *CoRR*, vol. abs/1605.04665, 2016. [Online]. Available: http://arxiv.org/abs/1605.04665

[4] Remark: Also the terminated ensemble can slightly benefit from optimized mapping as in [5, Fig. 8]. This is not considered here. However, rate loss occurs and thus the effective gap to capacity changes.